\documentclass[10pt,a4paper]{article}
\usepackage[T1]{fontenc}
\usepackage[utf8]{inputenc}
\usepackage[english]{babel}
\usepackage{lmodern}
\usepackage[a4paper,margin=18mm,columnsep=6mm]{geometry}
\usepackage{microtype}
\usepackage{amsmath,amssymb}
\usepackage{graphicx}
\usepackage{booktabs}
\usepackage{marvosym}
\usepackage[font=small,labelfont=bf,textfont=it,labelsep=period]{caption}
\usepackage[numbers,sort&compress]{natbib}
\usepackage[hidelinks]{hyperref}
\graphicspath{{figures/}}
\newcommand{\CdA}{C_{\mathrm d}A}
\newcommand{\Crr}{C_{\mathrm{rr}}}
\newcommand{\Pforty}{P_{40}}

\begin{document}
\twocolumn[{
\begin{center}
{\Large\bfseries\boldmath Quantifying Bicycle Setup Differences on the Road:\\
Field Estimates of CdA, Crr and Power Demand\par}
\vspace{7pt}
{\normalsize Mario Zinßer\par}
\vspace{3pt}
{\small tri2max, \Letter\,\href{mailto:info@tri2max.app}{info@tri2max.app}\par}
\vspace{8pt}
\begin{minipage}{0.96\textwidth}
\small
Individual bicycle configurations can be assessed in situ using power and speed measurements without a velodrome or wind tunnel. We apply an integrated energy-balance workflow to five configurations tested by one rider on a road section. Five nominal power levels, each ridden in both directions, yield 25 paired observations. Kinetic-energy changes are corrected explicitly; gravitational work cancels between spatially matched outward and return sections. A shared effective rolling-resistance coefficient $\Crr$ and a separate aerodynamic drag area $\CdA$ for each setup are estimated jointly. The fitted drag areas range from $0.2288$ to $0.3601\,\mathrm{m}^{2}$, with a common $\Crr=0.00551\pm0.00047$. At $40\,\mathrm{km/h}$, the estimated differences are $17.4\pm5.5\,\mathrm{W}$ favouring the EZ-covered over the uncovered $80\,\mathrm{mm}$ rear-wheel configuration and $64.3\pm4.0\,\mathrm{W}$ favouring the aerobar position over the drops on the tested road bike. The helmet swap yields a difference of $3.5\pm2.8\,\mathrm{W}$. Reported uncertainties combine statistical variability and assumed input uncertainties and represent one standard deviation. The study demonstrates a practical workflow for quantifying individual setup differences and connecting resistance estimates to power-speed predictions under specified reference conditions.
\end{minipage}
\end{center}
\vspace{12pt}
}]

\section{Introduction}
The aerodynamic performance of cycling equipment depends on its interaction with the rider and riding position. Testing the actual rider-bicycle system on the road provides an accessible way to assess individual configurations in their intended environment. Such in-situ measurements avoid the need for a separate laboratory, wind tunnel or velodrome.

The physical power balance is established \citep{Martin1998Power}. Field regression across different riding speeds \citep{Lim2011Resistance}, virtual-elevation methods \citep{Chung2012CdA} and inverse power-model fitting \citep{Danek2021Power} provide foundations for resistance estimation. Here, the contribution is a documented application to five practical configurations using paired road sections, integrated measured signals and jointly propagated setup comparisons. We investigate how the estimated resistance parameters translate into standardized power demand and whether the principal comparisons persist across processing and model variants.

\section{Physical foundations}
\subsection{Resistance and energy balance}
Let $m$ denote total system mass, $v$ ground speed, $\rho$ air density, $\theta$ road inclination and $w$ the signed headwind component (positive for headwind). The effective rolling-resistance coefficient is $\Crr$, and $\CdA$ is the product of drag coefficient and reference area. With gravitational acceleration $g=9.80665\,\mathrm{m/s^2}$, a one-dimensional model with positive relative air speed gives
\begin{align}
F_{\mathrm{roll}}&=\Crr mg\cos\theta,\\
F_{\mathrm{aero}}&=\tfrac12\rho\CdA(v+w)^2.
\end{align}
Mechanical power is force multiplied by ground speed. Thus aerodynamic power is cubic in $v$ in still air. For constant drivetrain efficiency $\eta$, without braking, the integrated balance is
\begin{align}
\eta E_p &= E_{\mathrm{roll}}+E_{\mathrm{aero}}
             +\Delta E_{\mathrm{kin}}+\Delta E_{\mathrm{pot}},\label{eq:energy}\\
E_p &= \int P_{\mathrm{pedal}}\,dt,\nonumber\\
E_{\mathrm{roll}}&=\Crr mg\int\cos\theta\,ds,\nonumber\\
E_{\mathrm{aero}}&=\tfrac12\rho\CdA\int(v+w)^2v\,dt,\nonumber\\
E_{\mathrm{kin}}&=\tfrac12mv^2+\tfrac12\sum_j I_j\omega_j^2,\nonumber\\
E_{\mathrm{pot}}&=mgh.\nonumber
\end{align}
Here, $P_{\mathrm{pedal}}$ is measured pedal power, $t$ time, $s$ distance travelled and $h$ elevation. For each stored-energy term, $\Delta E_x=E_x(t_1)-E_x(t_0)$ between section entry $t_0$ and exit $t_1$. The quantities $I_j$ and $\omega_j$ denote the moment of inertia and angular speed of rotating component $j$. The kinetic term includes translational and, when represented explicitly, rotational energy. The fitted $\Crr$ is an effective field-loss parameter rather than an isolated tire laboratory measurement.

\subsection{Paired sections, elevation and wind}
Gravitational work depends only on endpoint elevations. Consequently, $mg(h_B-h_A)+mg(h_A-h_B)=0$ for matched sections, irrespective of unequal speeds or durations. Changes in aerodynamic losses caused by the speed profile remain in the integral; averaging speed before cubing it would discard this information. Kinetic-energy changes must also be retained separately.

Wind does not generally cancel. With constant longitudinal wind and equal speeds in both directions,
\begin{equation}
\overline{P}_{\mathrm{aero}}
=\tfrac12\rho\CdA\left(v^3+w^2v\right).
\end{equation}
The linear wind contribution cancels in this symmetric case, but the quadratic term remains. Unequal speeds, crosswind and changing conditions require further treatment \citep{Polanco2024WindGrade}. Our primary fit uses a still-air approximation; wind is examined through sensitivity scenarios.

\section{Materials and methods}
\subsection{Configurations and field protocol}
\begin{figure}[!tbp]
\centering
\includegraphics[width=\columnwidth]{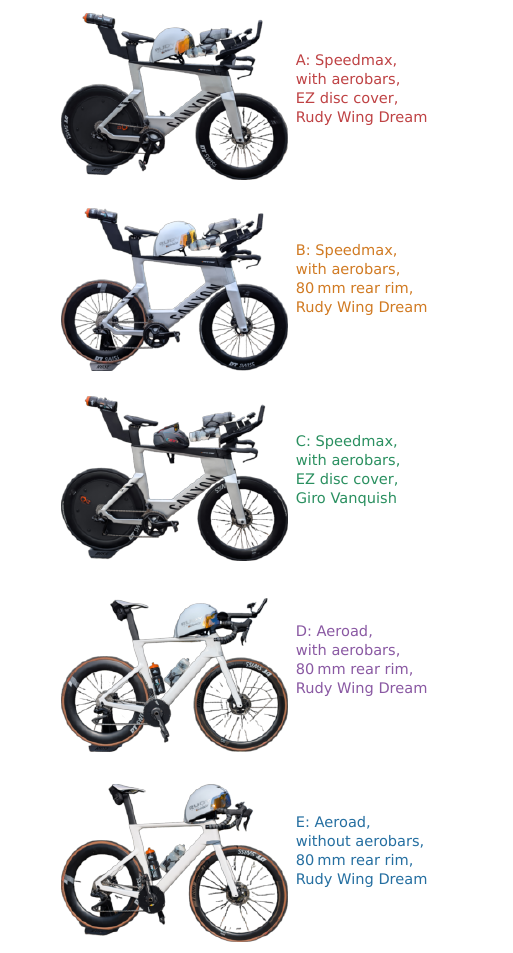}
\caption{The five tested configurations. Photos identify equipment; they do not show the rider's measured posture. Tire components and nominal pressures were identical.}
\label{fig:setups}
\end{figure}

The experiment was performed on 29 October 2026 near $47.47181^{\circ}\,\mathrm{N}$, $9.69185^{\circ}\,\mathrm{E}$. The five configurations are defined in Table~\ref{tab:setups} and Fig.~\ref{fig:setups}. Speedmax and Aeroad denote Canyon bicycle models; aerobars denote the forward extensions and arm supports used for the aero position. DT Swiss ARC 1100 wheels were used, with an applied EZ disc cover on the rear wheel in A/C and an uncovered $80\,\mathrm{mm}$ rear rim in B/D/E. All setups used Continental GP 5000 TT tires, nominally $25\,\mathrm{mm}$ front and $28\,\mathrm{mm}$ rear, with Eclipse Road Endurance thermoplastic polyurethane (TPU) tubes, model ECRD3540. Tire pressure was $6\,\mathrm{bar}$ front and rear. The rider remained on the aerobars in A-D and rode on the drops in E. The front wheel, clothing, shoes, bottles and other accessories were constant. The tested order was E, D, C, A, B.

\begin{table*}[t]
\centering
\caption{Tested configurations and total rider-bicycle-equipment masses. Rudy denotes Rudy Project Wing Dream; Giro denotes Giro Vanquish. Disc denotes a rear wheel fitted with an EZ disc cover, not a dedicated disc wheel. The $80\,\mathrm{mm}$ dimension is rear-rim depth. Tire and tube components were identical, with $6\,\mathrm{bar}$ front and rear.}
\label{tab:setups}
\small
\setlength{\tabcolsep}{8pt}
\renewcommand{\arraystretch}{1.25}
\begin{tabular}{@{}lp{0.62\textwidth}r@{}}
\toprule
Setup & Configuration & Mass / $\mathrm{kg}$ \\
\midrule
A & Speedmax, with aerobars, EZ disc cover, Rudy & 93.95 \\
B & Speedmax, with aerobars, $80\,\mathrm{mm}$ rear rim, Rudy & 93.30 \\
C & Speedmax, with aerobars, EZ disc cover, Giro & 93.75 \\
D & Aeroad, with aerobars, $80\,\mathrm{mm}$ rear rim, Rudy & 92.70 \\
E & Aeroad, without aerobars, $80\,\mathrm{mm}$ rear rim, Rudy & 92.30 \\
\bottomrule
\end{tabular}
\end{table*}

Nominal levels of $80$, $150$, $220$, $290$ and $360\,\mathrm{W}$ were each ridden outward and back. Power was measured with the same dual-sided Favero Assioma pedal sensors in all setups, and speed with a Garmin Speed Sensor 2. Power, speed and position data were recorded at $1\,\mathrm{Hz}$ on a Garmin Fenix 8. The measured power, rather than the assigned level, enters the fit. Ambient conditions were represented by one constant air density per configuration, calculated as an ideal-gas mixture from the setup-mean temperature, humidity and surface pressure of archived Open-Meteo weather-model data \citep{OpenMeteo2026Weather}. These are model estimates rather than on-site readings.

\subsection{Processing and joint estimation}
\begin{figure}[!tbp]
\centering
\includegraphics[width=\columnwidth]{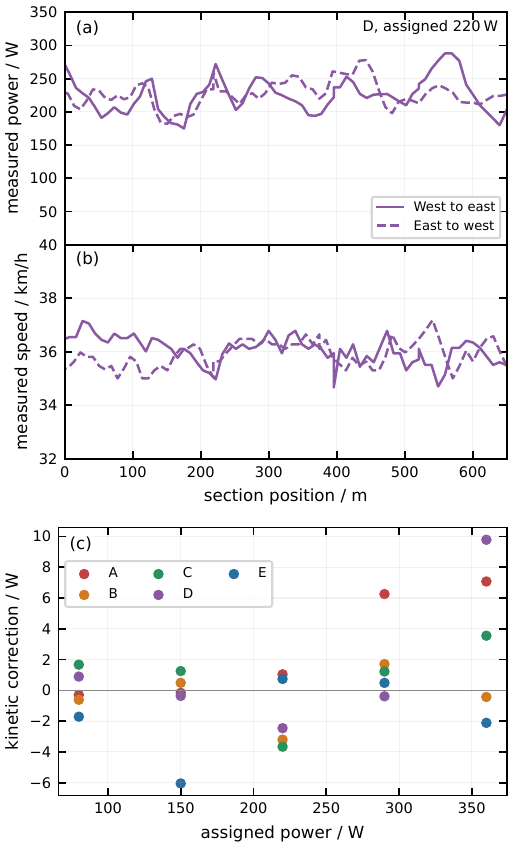}
\caption{Measured power and speed for a spatially matched outward/return pair (a,b), and kinetic corrections for all pairs (c). The correction is the paired kinetic-energy change divided by $\eta$ and total pair duration, subtracted from mean pedal power. The common position axis is a translation of the internal crop coordinate.}
\label{fig:processing}
\end{figure}

The ten measurement laps per setup were paired by assigned level and direction. A common spatial window, $150$--$800\,\mathrm{m}$ on the internal route axis, yielded a $650\,\mathrm{m}$ analysis section. Its plotted coordinate runs from zero at the western boundary to $650\,\mathrm{m}$ at the eastern boundary. Figure~\ref{fig:processing} illustrates the processing using setup D at the assigned $220\,\mathrm{W}$ level, chosen for a calmer power trace; all 25 pairs remain in the analysis.

Boundary samples were interpolated in time. Integrals of power, speed and speed cubed were evaluated from piecewise-linear signals. Endpoint speeds were estimated with local linear fits over approximately $5\,\mathrm{s}$. The primary kinetic correction used translational energy; rotational inertia was checked separately. With $\cos\theta\simeq1$, matched endpoints and $w=0$, each pair supplies
\begin{equation}
\begin{split}
\eta\sum_\ell E_{p,\ell}-\sum_\ell\Delta E_{\mathrm{kin},\ell}
={}&\Crr mg\sum_\ell\int v_\ell\,dt\\
&+\tfrac12\rho\CdA\sum_\ell\int v_\ell^3\,dt,
\end{split}\label{eq:pair}
\end{equation}
where $\ell$ indexes the two directions. We divided each balance by total pair duration and fitted one common $\Crr$ and five setup-specific $\CdA$ values by ordinary least squares, with equal weight per pair. There are 25 observations, six coefficients and 19 residual degrees of freedom. Drivetrain efficiency was fixed at $\eta=0.975$. One, two and five rolling-resistance parameters were compared; the two-coefficient alternative groups A/C and B/D/E by rear-wheel configuration.

\subsection{Uncertainty and reference predictions}
\begin{figure}[!tbp]
\centering
\includegraphics[width=\columnwidth]{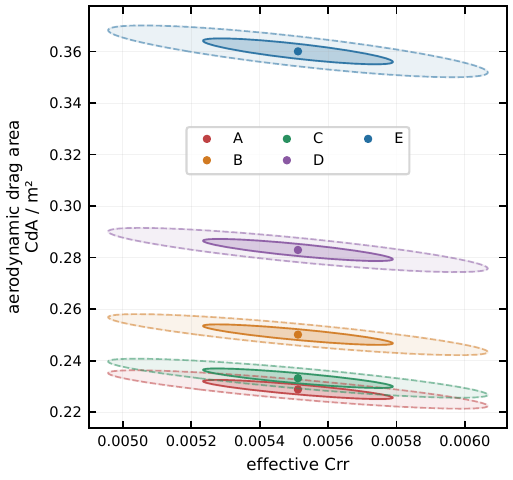}
\caption{Statistical coupling of common $\Crr$ and setup-specific $\CdA$. Points are nominal fits. Dark inner and light outer regions have covariance-scaled radii one and two; outer contours are dashed. These Mahalanobis contours are not marginal error bars or joint 68/95-percent confidence regions. The additional input budget is excluded here.}
\label{fig:parameter-correlation}
\end{figure}

Parameter coupling is represented by the statistical covariance ellipses in Fig.~\ref{fig:parameter-correlation}. Combined uncertainty was estimated using 30,000 independent draws of input perturbations and statistical parameter deviations. For the latter, let $X$ be the design matrix of the time-normalized pair balances and $s^2$ their residual variance estimate with $\nu=19$ degrees of freedom. We used
\[
\begin{gathered}
\delta\boldsymbol\beta=\boldsymbol z\sqrt{\nu/u},\quad u\sim\chi^2_\nu,\\
\boldsymbol z\sim\mathcal N(\boldsymbol0,C),\quad
C=s^2(X^{\mathsf T}X)^{-1}.
\end{gathered}
\]
The variables $\boldsymbol z$ and $u$ are independent. Thus $C$ is the Student-$t$ scale matrix; the statistical covariance is $19C/17$. This component assumes independent Gaussian pair errors with common variance. Each combined parameter draw is the input-perturbed refit plus an independent $\delta\boldsymbol\beta$. Monte Carlo propagation provides the framework \citep{JCGM2008GUM,JCGM2008MonteCarlo}; the additive combination is an approximation.

Input perturbations were independent between error types and normally distributed about their nominal values. Assumed standard deviations were a power gain of $1\,\%$ and offset of $2\,\mathrm{W}$ shared by all setups, plus independent $1\,\mathrm{W}$ setup offsets; a $1\,\%$ density scale shared by all setups; independent $0.1\,\mathrm{kg}$ setup masses; and two independent speed-scale errors of $0.5\,\%$, shared within A/C and B/D/E respectively. The speed grouping is a propagation assumption. Efficiency draws had mean $0.975$ and standard deviation $0.005$, shared within A/B/C and D/E respectively, and were clipped to $[0.94,1]$. These are engineering assumptions, not calibration records.

For each input draw, power means, the linear and cubic speed integrals, and kinetic corrections were rescaled before refitting; section boundaries and durations were held fixed. The kinetic term scales with mass and speed squared. Draws outside $0<\Crr<0.03$ or $0<\CdA<1\,\mathrm{m^2}$ were rejected. Reference predictions used nominal masses and fixed reference density with each draw's fitted parameters and bicycle efficiency; perturbed masses and densities entered through the refits. Reported uncertainties are the standard deviations of the retained draws about their draw means, displayed around the nominal fits. Differences were calculated within the same joint draws to preserve shared errors and parameter correlations. Unmeasured wind is addressed separately by sensitivity scenarios.

Reference curves and the predicted pedal power at $40\,\mathrm{km/h}$, denoted $\Pforty$, assume level road, still air, each setup's mass, $\eta=0.975$ and common density $\rho_{\mathrm{ref}}=1.1905\,\mathrm{kg/m^3}$. Robustness checks used alternative windows of $200$--$650\,\mathrm{m}$ and $200$--$700\,\mathrm{m}$ on the internal axis and varied endpoint estimation, the $\Crr$ structure, power timing, rotational mass and residual weighting. Internal leave-one-pair-out predictions were also evaluated. None of these checks is an independent reference measurement.

\section{Results}
\subsection{Resistance and standardized performance}
The common-$\Crr$ model gives $\Crr=0.00551\pm0.00047$. The negative ellipse slopes in Fig.~\ref{fig:parameter-correlation} show the finite experimental separation of rolling and aerodynamic losses: a higher rolling estimate can be offset partly by a lower drag area. Joint estimation preserves this relationship when deriving performance predictions.

\begin{figure}[!tbp]
\centering
\includegraphics[width=\columnwidth]{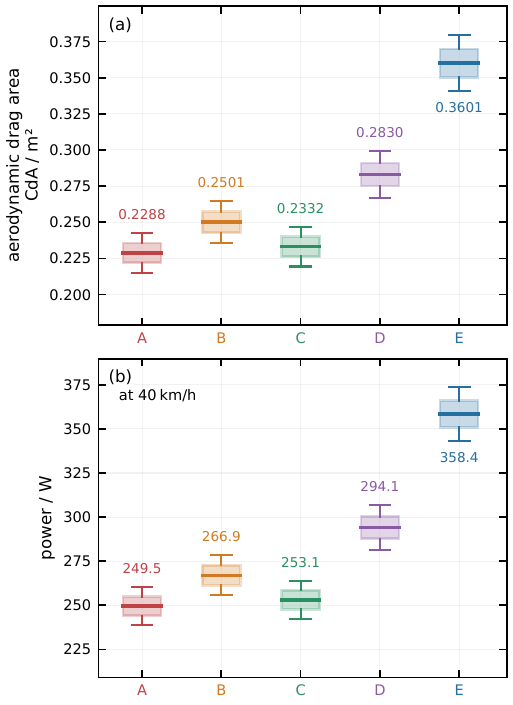}
\caption{Drag area and reference power demand. Center lines and numeric labels show nominal fits; box edges show fit $\pm1\sigma$, whiskers fit $\pm2\sigma$. These are defined uncertainty boxes, not quartile boxplots; the whiskers are not exact 95-percent intervals.}
\label{fig:resistance-results}
\end{figure}

Table~\ref{tab:results} and Fig.~\ref{fig:resistance-results} report the configuration estimates. Setup A has the lowest nominal drag area and reference power demand. Across setups, estimated $\CdA$ ranges from $0.2288$ to $0.3601\,\mathrm{m^2}$ and $\Pforty$ from $249.5$ to $358.4\,\mathrm{W}$.

\begin{table*}[t]
\centering
\caption{Nominal estimates with $\pm1\sigma$ combined statistical and assumed input uncertainty. The rolling coefficient is one shared parameter. Power assumes level road, still air, setup-specific masses, $\rho=1.1905\,\mathrm{kg/m^3}$ and $\eta=0.975$.}
\label{tab:results}
\normalsize
\setlength{\tabcolsep}{9pt}
\renewcommand{\arraystretch}{1.3}
\begin{tabular}{@{}lccc@{}}
\toprule
Setup & $\CdA$ / $\mathrm{m^2}$ & $\Crr$ & $\Pforty$ / $\mathrm{W}$ \\
\midrule
A & $0.2288\pm0.0068$ & $0.00551\pm0.00047$ & $249.5\pm5.4$ \\
B & $0.2501\pm0.0073$ & $0.00551\pm0.00047$ & $266.9\pm5.7$ \\
C & $0.2332\pm0.0069$ & $0.00551\pm0.00047$ & $253.1\pm5.4$ \\
D & $0.2830\pm0.0080$ & $0.00551\pm0.00047$ & $294.1\pm6.3$ \\
E & $0.3601\pm0.0098$ & $0.00551\pm0.00047$ & $358.4\pm7.6$ \\
\bottomrule
\end{tabular}
\end{table*}

\begin{figure}[!tbp]
\centering
\includegraphics[width=\columnwidth]{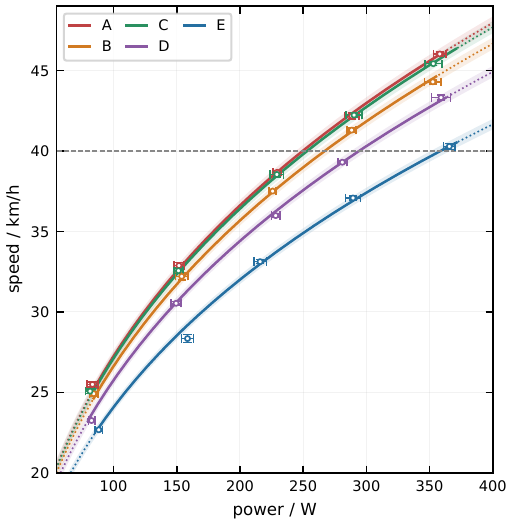}
\caption{Speed-power curves with $\pm1\sigma$ combined power-uncertainty bands at fixed speed and paired data with horizontal and vertical $\pm1\sigma$ bootstrap bars. Point bars exclude calibration, endpoint and weather-drift uncertainty. Dotted curves extend beyond the observed range of section-mean speeds.}
\label{fig:speed-power}
\end{figure}

Figure~\ref{fig:speed-power} translates the estimates into speed-power curves. Measured markers use kinetic-corrected pair power and cubic-mean equivalent speed,
\[
v_{\mathrm{eq}}=
\left(\frac{\sum_\ell\int v_\ell^3\,dt}{\sum_\ell T_\ell}\right)^{1/3}.
\]
Here, $T_\ell$ is the duration of direction $\ell$. This equivalent speed preserves the cubic aerodynamic term; rolling work uses the separately integrated speed. The points represent measurements under each setup's ambient conditions, whereas the curves use the common reference density; the points are not adjusted using fitted resistance parameters. Point errors were estimated by a within-pass circular block bootstrap with $10\,\mathrm{s}$ blocks and 4,000 draws, resampling power and speed jointly. They describe within-pass sampling variability rather than repeated-test precision \citep{Hopkins2000Reliability}.

\subsection{Configuration differences}
\begin{figure}[!tbp]
\centering
\includegraphics[width=\columnwidth]{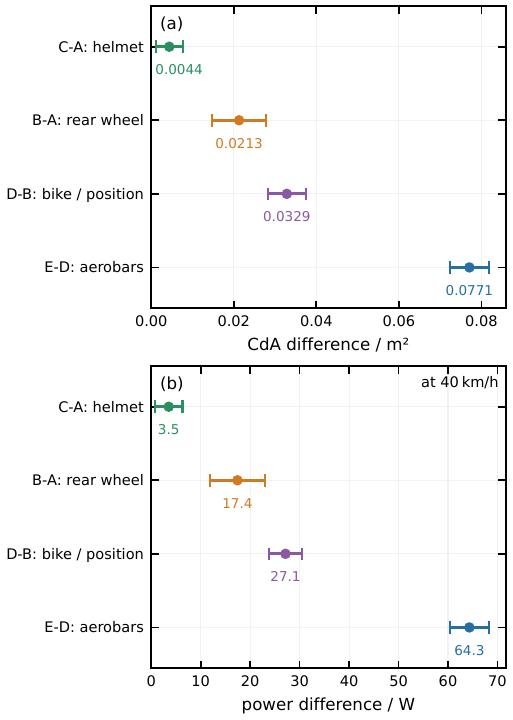}
\caption{Nominal differences with $\pm1\sigma$ joint uncertainty. Positive values indicate a larger requirement for the first setup: $\mathrm{C}-\mathrm{A}$ helmet, $\mathrm{B}-\mathrm{A}$ uncovered $80\,\mathrm{mm}$ versus EZ-covered rear-wheel configuration, $\mathrm{D}-\mathrm{B}$ bicycle/riding configuration, $\mathrm{E}-\mathrm{D}$ drops without aerobars versus aerobar position. Numeric labels are nominal differences rather than draw means.}
\label{fig:setup-differences}
\end{figure}

All differences subtract the second named setup from the first; positive $\Delta\Pforty$ indicates higher predicted pedal power. Comparing the uncovered $80\,\mathrm{mm}$ rear-wheel configuration with the EZ-covered configuration gives $\Delta\Pforty=17.4\pm5.5\,\mathrm{W}$ for B minus A, favouring the EZ-covered configuration. For the tested configurations with aerobars, D minus B gives $27.1\pm3.2\,\mathrm{W}$, favouring the Speedmax configuration over the Aeroad configuration. Comparing the drops position without aerobars with the aerobar position on the Aeroad gives E minus D of $64.3\pm4.0\,\mathrm{W}$; this contrast includes both position and equipment changes. The Giro-minus-Rudy helmet comparison, C minus A, yields a smaller $3.5\pm2.8\,\mathrm{W}$ estimate favouring Rudy. Figure~\ref{fig:setup-differences} shows both drag-area and power differences with their jointly propagated uncertainty.

\subsection{Model fit and robustness}
The primary model has residual standard deviation $3.59\,\mathrm{W}$ and internal leave-one-pair-out root-mean-square error (RMSE) $4.07\,\mathrm{W}$. Relative values of the small-sample corrected Akaike information criterion (AICc, lower is better) for one, two and five rolling coefficients are $0$, $0.54$ and $8.55$. The common coefficient therefore provides a parsimonious primary model consistent with the identical tire components.

\begin{figure}[!tbp]
\centering
\includegraphics[width=\columnwidth]{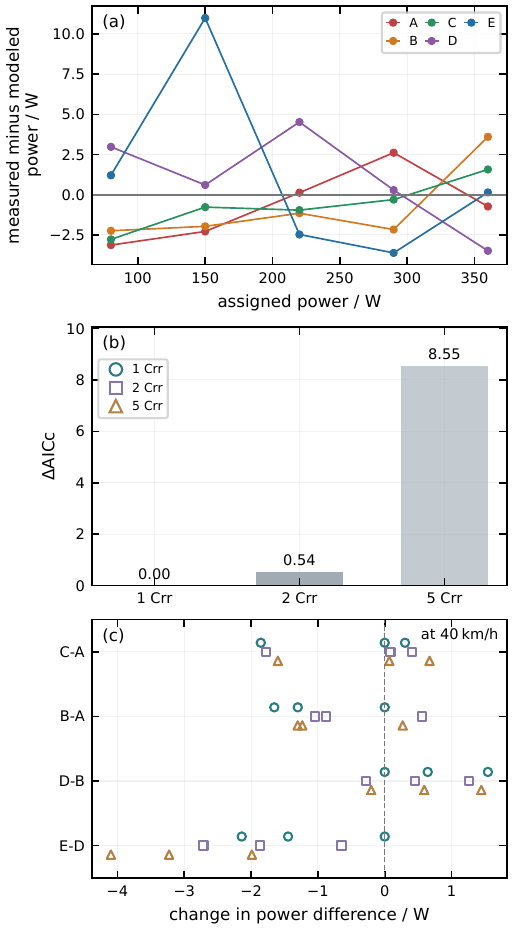}
\caption{Pair residuals (a), $\Crr$-model comparison using the corrected Akaike information criterion (AICc; b), and changes in the four power differences relative to the primary fit (c). Panel (c) includes all nine combinations of three crops and one, two or five rolling coefficients, with $5\,\mathrm{s}$ endpoint estimation. Sensitivity points are scenarios, not additional statistical error bars.}
\label{fig:model-robustness}
\end{figure}

Figure~\ref{fig:model-robustness} shows all nine combinations of three crop windows and three $\Crr$ structures using the primary endpoint treatment. The larger configuration differences persist across these variants, while the allocation to rolling and aerodynamic parameters shifts, consistent with Fig.~\ref{fig:parameter-correlation}. Shifting power timestamps by $\pm1\,\mathrm{s}$, adding $3\,\mathrm{kg}$ of rotationally equivalent mass, or using the tested robust fit each changed setup $\Pforty$ by less than $1\,\mathrm{W}$. Using a leverage-adjusted, heteroskedasticity-consistent covariance estimator (HC3) to allow unequal residual variances changed the combined drag-area standard deviations by factors of approximately $0.97$--$1.16$.

\section{Discussion}
The experiment demonstrates a practical road-based route from recorded power and speed to individual equipment and position comparisons. Integrated signals allow nonconstant speed to be accommodated explicitly, while paired spatial boundaries remove net gravitational work. The result is a physically interpretable comparison without a dedicated test facility, building on established field methods \citep{Lim2011Resistance,Chung2012CdA,Danek2021Power}.

Setup A achieved a low estimated drag area of $0.2288\pm0.0068\,\mathrm{m^2}$ for a rider measuring $192\,\mathrm{cm}$ and weighing $80.5\,\mathrm{kg}$. The estimate is consistent with published measurements in time-trial positions \citep{Faulkner2024Individual} and indicates an aerodynamically efficient rider-bicycle configuration.

The largest observed configuration changes are clearly separated relative to their reported uncertainty. The rear-wheel configuration comparison also yields a substantial nominal reduction in reference power. The helmet estimate illustrates the finer end of the comparison: its direction favours Rudy, but its size is comparable to the uncertainty, so it is reported as an estimated effect rather than a generally superior helmet. The $\mathrm{D}-\mathrm{B}$ comparison includes the positions actually ridden and is not a frame-only test. Equipment conclusions apply to this rider and these configurations.

The main conditions for interpretation concern environmental and input assumptions. The tests were completed in successive blocks rather than repeated randomized comparisons; temporal changes can therefore accompany setup changes. Local wind was not measured. In ideal symmetric riding, its remaining quadratic contribution has the same speed dependence as rolling power, illustrating why small fit residuals alone cannot establish absolute parameter accuracy. Constant longitudinal-wind scenarios of $\pm2\,\mathrm{m/s}$ reduced the inferred still-air $\Pforty$ by approximately $5$--$10\,\mathrm{W}$ depending on setup and direction; these scenarios are separate from the reported $\sigma$ budget, not measured corrections.

The useful next validation step is repetition of matched comparison blocks, preferably with local wind observations and documented sensor calibration. This would test complete-protocol repeatability beyond the within-pass bootstrap. For the present data, transparent joint uncertainty and processing sensitivity support the practical comparisons while retaining the distinction between model fit, repeatability and reference accuracy.

\section{Conclusion}
Paired road measurements combined with an integrated energy balance quantified individual bicycle-configuration differences without a velodrome or wind tunnel. A common effective rolling coefficient and separate drag areas linked the measured signals to speed-power curves and reference power demand. The field workflow resolved substantial wheel and riding-configuration differences and produced a smaller, uncertainty-qualified helmet estimate. It provides an accessible basis for individual setup assessment under documented conditions.

\paragraph{AI-assisted writing.}
AI-assisted tools were used to refine the wording of this manuscript. The author reviewed the text and takes responsibility for its scientific content.

\bibliographystyle{unsrtnat}
\bibliography{references}

@article{Faulkner2024Individual,
  title = {{Individual aerodynamic and physiological data are critical to optimise cycling time trial performance: one size does not fit all}},
  year = {2024},
  url = {https://link.springer.com/article/10.1007/s12283-023-00446-0},
  author = {Steve H. Faulkner and Philippa Jobling and Katy E. Griggs and Petros Siegkas},
  doi = {10.1007/s12283-023-00446-0},
  journal = {Sports Engineering},
  volume = {27},
  eid = {4},
  note = {Article 4}
}

@misc{Chung2012CdA,
  title = {{Estimating CdA with a power meter}},
  year = {2012},
  url = {https://raw.githubusercontent.com/GoldenCheetah/GoldenCheetah/master/doc/contrib/ChungVE.pdf},
  urldate = {2026-10-30},
  author = {Robert Chung},
  note = {Original presentation, version March 2012; retrieved from the GoldenCheetah repository.}
}

@misc{Danek2021Power,
  title = {{On modelling bicycle power-meter measurements}},
  year = {2021},
  url = {https://arxiv.org/html/2103.09806v1},
  urldate = {2026-10-30},
  author = {Tomasz Danek and Michael A. Slawinski and Theodore Stanoev},
  eprint = {2103.09806},
  archivePrefix = {arXiv},
  note = {arXiv:2103.09806, version 1.}
}

@article{Hopkins2000Reliability,
  title = {{Measures of reliability in sports medicine and science}},
  year = {2000},
  url = {https://pubmed.ncbi.nlm.nih.gov/10907753/},
  urldate = {2026-10-30},
  author = {Will G. Hopkins},
  doi = {10.2165/00007256-200030010-00001},
  journal = {Sports Medicine},
  volume = {30},
  number = {1},
  pages = {1-15}
}

@misc{JCGM2008GUM,
  title = {{Evaluation of measurement data - Guide to the expression of uncertainty in measurement}},
  year = {2008},
  url = {https://www.bipm.org/documents/20126/2071204/JCGM_100_2008_E.pdf},
  urldate = {2026-10-30},
  author = {{Joint Committee for Guides in Metrology}},
  doi = {10.59161/JCGM100-2008E}
}

@misc{JCGM2008MonteCarlo,
  title = {{Evaluation of measurement data - Supplement 1 to the Guide to the expression of uncertainty in measurement - Propagation of distributions using a Monte Carlo method}},
  year = {2008},
  url = {https://www.bipm.org/en/web/guest/publications/guides},
  urldate = {2026-10-30},
  author = {{Joint Committee for Guides in Metrology}},
  doi = {10.59161/JCGM101-2008}
}

@article{Lim2011Resistance,
  title = {{Measuring changes in aerodynamic/rolling resistances by cycle-mounted power meters}},
  year = {2011},
  url = {https://pubmed.ncbi.nlm.nih.gov/20881880/},
  urldate = {2026-10-30},
  author = {Allen C. Lim and Eric P. Homestead and Andrew G. Edwards and Todd C. Carver and Rodger Kram and William C. Byrnes},
  doi = {10.1249/MSS.0b013e3181fcb140},
  journal = {Medicine \& Science in Sports \& Exercise},
  volume = {43},
  number = {5},
  pages = {853-860}
}

@article{Martin1998Power,
  title = {{Validation of a Mathematical Model for Road Cycling Power}},
  year = {1998},
  url = {https://collections.lib.utah.edu/dl_files/b4/8e/b48ef26086091662c561e673d7bd990d77868437.pdf},
  urldate = {2026-10-30},
  author = {James C. Martin and Douglas L. Milliken and John E. Cobb and Kevin L. McFadden and Andrew R. Coggan},
  doi = {10.1123/jab.14.3.276},
  journal = {Journal of Applied Biomechanics},
  volume = {14},
  number = {3},
  pages = {276-291}
}

@misc{OpenMeteo2026Weather,
  author = {{Open-Meteo}},
  title = {{Historical Weather API documentation}},
  year = {2026},
  url = {https://open-meteo.com/en/docs/historical-weather-api},
  note = {Accessed 2 November 2026}
}

@article{Polanco2024WindGrade,
  title = {{Influence of wind speed and road grade on the estimation of drag area in cycling}},
  year = {2024},
  url = {https://pubmed.ncbi.nlm.nih.gov/33433293/},
  urldate = {2026-10-30},
  author = {Alejandra Polanco and Sergio Roa and Daniel R. Su{\'a}rez and Omar L{\'o}pez and Luis Mu{\~n}oz},
  doi = {10.1080/14763141.2020.1837925},
  journal = {Sports Biomechanics},
  volume = {23},
  number = {2},
  pages = {222-240}
}
\end{document}